# Nanotube-Terminated Zigzag Edge of Phosphorene formed by Self-Rolling Reconstruction

Junfeng Gao, Xiangjun Liu, Gang Zhang*, and Yong-Wei Zhang+

*Institute of High Performance Computing, A*STAR, Singapore, 138632, Singapore.*

Edge atomic configuration often plays an important role in dictating the properties of finite-sized two-dimensional (2D) materials. By performing *ab initio* calculations, we identify a highly stable zigzag edge of phosphorene, which is the most stable one among all the considered edges. Surprisingly, this highly stable edge exhibits a novel nanotube-like structure, which is topologically distinctively different from any previously reported edge reconstruction. We further show that this new edge type can form easily, with an energy barrier of only 0.234 eV. It may be the dominant edge type at room temperature in vacuum condition or even under low hydrogen gas pressure. The calculated band structure reveals that the reconstructed edge possesses a bandgap of 1.23 eV. It is expected that this newly found edge structure may stimulate more studies in uncovering other novel edge types and further exploring their practical applications.

Authors' information

J. Gao : gaojf@ihpc.a-star.edu.sg

X. Liu: liux@ihpc.a-star.edu.sg

G. Zhang: zhangg@ihpc.a-star.edu.sg

Y.-W. Zhang: zhangyw@ihpc.a-star.edu.sg

**Introduction**

Phosphorene, a single-layer of black phosphorus, which possesses a puckered honeycomb lattice, has recently experienced a surge of research interest.[1, 2] Since its lattice structure is highly anisotropic, its mechanical, thermal and electronic properties are also highly anisotropic.[3, 4] In particular, it was shown[5] that phosphorene is a semiconductor with a direct band gap of ~2.0 eV, and exhibits an on-off current ratio above $10^5$. In addition, the electronic properties of black phosphorus can be efficiently tuned via the number of layers, strain, vacancy, and electronic field.[6-8] Evidently, phosphorene possesses many remarkable properties ideal for electronic and optoelectronic device applications.

Just like a surface is an integral part of any finite-size three-dimensional materials, an edge is an integral part of a finite-size two-dimensional (2D) material.[9-13] It was shown that edge states in graphene nanoribbons exhibit electronic properties distinctively different from the bulk states,[14-16] such as spin-dependent gapless chiral edge states[14] and pseudo-Landau levels[15]. Similarly, edge states can also have strong effect on the electronic properties of phosphorene. For example, previous theoretical studies revealed that phosphorene nanoribbons (PNRs) with pristine armchair (ac) edges are semiconducting,[17, 18] while PNRs with pristine zigzag (zz) edges are metallic[19-21]. Upon edge hydrogenation, both zz and ac PNRs are semiconductors; while the band gap vs. width (w) follows 1/w relation for the former and a $1/w^2$ relation for the latter, bringing different optical responses[22]. It was also reported that PNRs exhibit a giant Stark effect and high Seebeck coefficient, with the former enabling the field-effect transistor to work under a low bias[23] and the latter being important for thermoelectric application[24]. Moreover, an antiferromagnetic insulating state was revealed for zz edge of PNRs, which may find new applications in nanoelectronics[25].

In general, a pristine edge with unsaturated dangling bonds is unstable, leading to the atomic reconstruction of the edge. In 2D honeycomb lattices, due to the short distance of unsaturated dangling bonds at ac edges, a simple edge reconstruction by forming triple bonds in the armrests can easily lower the edge energy.[9, 10] For zz edges, however, the distance between two dangling bonds are too far to form any triple bonds. Therefore, to enhance the stability of zz edges, a complex reconstruction is often required. For example, by transforming two hexagons into a pentagon and a heptagon, pristine graphene zz edge can transform into the well-known Haeckelite zz(57) edge, reducing the energy by ~2.1 eV/nm.[9] For transition metal dichalcogenides (TMDs), metal-terminated edges preferably undergo a unique (2×1) reconstruction by pushing half second-row halogens atoms outwards, causing, for example, an energy reduction of ~0.4 eV/nm for $MoS_2$, ~1.0 eV/nm for $MoSe_2$, ~1.2 eV/nm for $WS_2$ and ~1.4 eV/nm for $WSe_2$.[26] Compared to the extensive studies on the reconstruction of zz edges in graphene and TMDs,[9-12, 26] the study on the reconstruction of phosphorene zz edges remains rather limited. Due to the difficulties in directly imaging the edge structures experimentally, how a pristine phosphorene zz edge reconstructs itself remains unknown.

In this paper, we systemically study the stabilities and probable reconstructions of both zz and zz(Klein)[27] (zz(K)-type) edges, and their influence on the electronic properties via density functional theory (DFT) calculations. Surprisingly, we reveal a new form of 2D edge that is terminated by a nanotube structure by an easily self-rolling process. Compared with pristine zz edge, ***the nanotube-terminated edge is able to reduce the edge energy by 35%***. Importantly, it possesses the lowest edge energy among all the edge types explored, suggesting that this unique edge type may be present in reality. We further discuss its effect on the electronic band structure. It is expected that the finding of this new form of edge structure may also inspire more efforts to look for other novel forms of edge structures, and their applications.

**Theory and computational methods**

All first-principles calculations were performed by the VASP.[28] The PBE functional[29] and PAW method[30] were used to describe the exchange-correlation functional and core electrons, respectively. DFT-D3 correction of Grimme was used to calculate the vdW interaction.[31] The kinetic energy cutoff was taken as 400 eV and the k-mesh was carefully tested (See the Fig. S1 in electronic supplementary information, ESI). The force criterion for structure optimization and climbing image nudged elastic band (cNEB) calculation[32] were taken both as 0.02 eV/Å. A very dense (20×1×1) k-mesh was used for the band structure calculations and STM simulations. In the present work, we employ a unit cell containing two periods along the edge for studying the edge reconstruction. Here, we focus on both pristine zz and zz(K)-type edges and their various reconstructed edge structures.

**Results and discussion**

**Tube-terminated edge reconstruction**

Fig. 1a shows the reconstruction of pristine zz edge. It is seen that a spontaneous (2×1) reconstruction (z1 configuration) for pristine zz edge occurs first. In this process, the neighbouring edge atoms (labelled by 1 and 2) undergo a 0.37 Å buckling in height and 0.21 Å shifting vertical to the edge. The edge atoms (red) of z1 are bent upwards with a displacement of 0.37~0.74 Å with reference to the inner P atoms. Subsequently, a series of edge conformation changes take place through bond bending and rolling (see the z1 to z4 configurations in Fig. 1a). Eventually, a novel edge structure terminated by a nanotube (the z5 configuration) is formed.

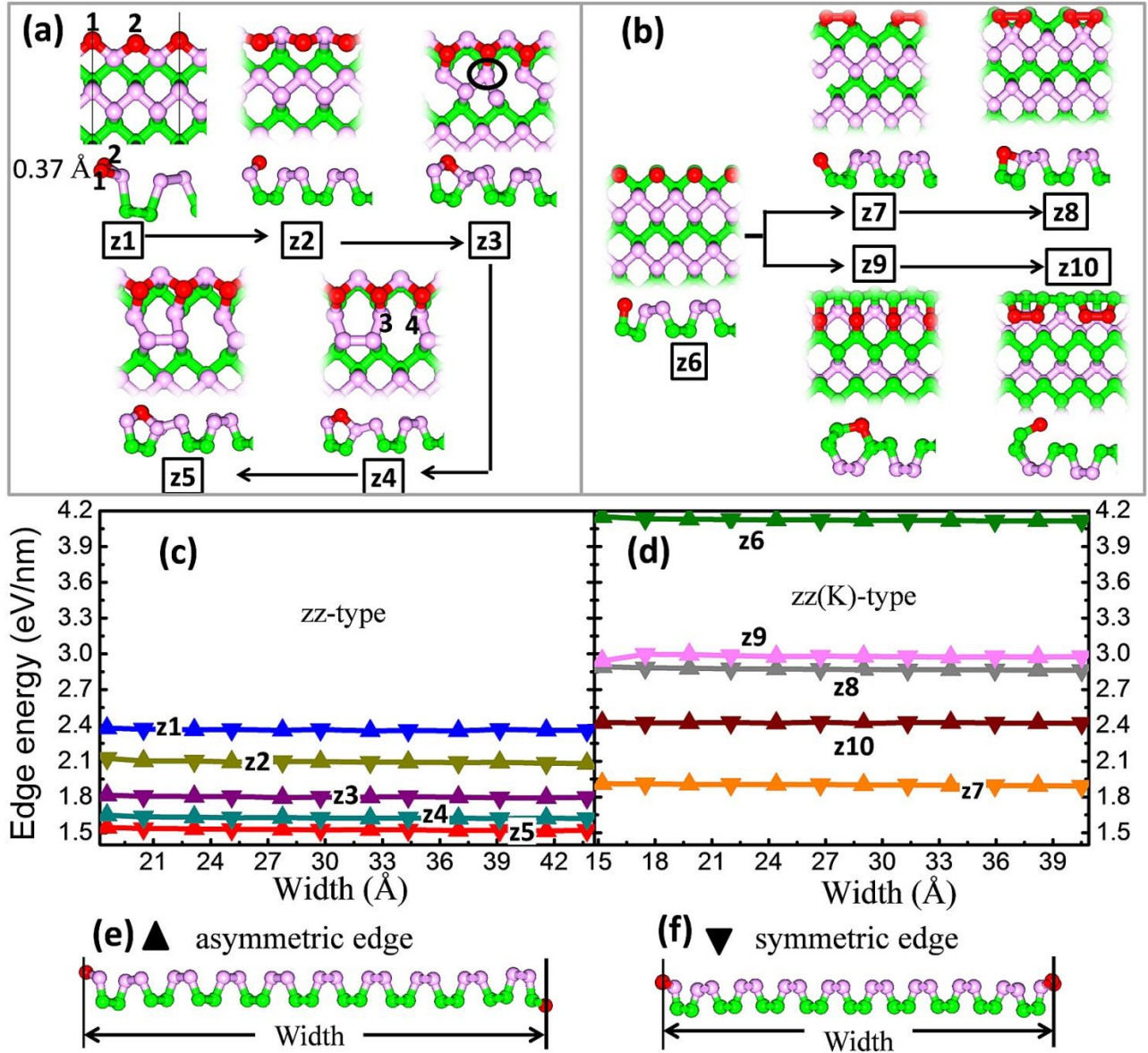

**Fig. 1** Possible zz-type (a) and zz(K)-type (b) edge structures and their edge energies (c)-(d), respectively. Side view of asymmetric (e) and symmetric (f) PNRs with the pristine z1 configuration. The definition of the width of PNR is also shown in (e) and (f).

To quantitatively understand the energetics during the above edge reconstruction of pristine zz edge, we calculated their edge energy per length γ using[9, 12]:

$$\gamma = \frac{1}{2L}(E_{PNR} - N_P E_P) \quad (1)$$

where $E_{PNR}$ is the total energy of a PNR, $N_P$ and $E_P$ are the number of P atoms in the PNR and the energy of P atom in perfect phosphorene, respectively. $L$ is the length of the PNR and the factor 2 accounts for the two edges of the PNR. Due to the puckered structure, a PNR can have two edges either in asymmetry (Fig. 1e) or symmetry (Fig. 1f). To exclude the interaction of two edges and achieve accurate edge energies, we have calculated the edge energy of a PNR with consecutively increasing its width over 43 Å. Here, the widths are defined as the width of pristine zz edge before reconstruction (Fig. 1e, 1f).

Fig. 1c shows the edge energy versus the PNR width, in which triangles (inverted triangles) represent the asymmetric edges (symmetric edge). The edge energy of the widest PNRs for each edge type is listed in Table 1. It is seen that the edge energy is neither dependent on the width, nor the nature of edge symmetry. The edge energy of the (2×1) reconstructed z1 configuration is 2.36 eV/nm, which is slightly lower than pristine zz edge (~0.08 eV/nm). Simply through flipping, the outward z1 configuration is able to transform into the inward z2 configuration. This process reduces the edge energy by 0.28 eV/nm (~11.8%). A further bending motion is able to connect the edge atom to the inner atoms in the same layer, and transform the z2 configuration into the z3 configuration. Since the z3 configuration eliminates the unsaturated bonds, its edge energy is reduced to only 1.80 eV/nm, giving rise to a 23.7% reduction.

It is noted that there are 4-coordinated P atoms (labelled by the black circle in Fig. 1a) in the z3 configuration. Upon a reallocation of P-P bonds, a higher-symmetry z4 configuration with all 3-coordinated P atoms is formed. In this process, the edge energy is reduced to ~1.62 eV/nm. The energy of z4 configuration can be further reduced by changing the orientation of the opposite lone pair electrons of atom-3 and atom-4 to eliminate the repelling interaction between them. This process transforms the z4 configuration into the nanotube-terminated z5 configuration. The edge energy of the z5 configuration is 1.52 eV/nm, which is only 64.4% of the z1 configuration. Clearly, among all the edges in the reconstruction process, the nanotube terminated z5 configuration possesses the lowest edge energy. This 1D nanotube and 2D atomic film combing system may exhibit novel properties like previous proposed graphene nanoribbon-nanotube junctions,[33] but it is noted that the phosphorene nanotubed edge can form by self-reconstruction with low transition barrier, which will be detailed discussed later. In principle, the two edge atoms can form bonds with phosphorous atoms in the inner positions, which results in tube-liked edge with a larger diameter. However, in this process, the flat phosphorene sheet needs to bend up, thus there is a drastic increase in the total energy. Based on this consideration, formation of tube-liked edge with a larger diameter is not energetically favourable.

Fig. 1b (1d) shows the structures (edge energy) of pristine and reconstructed zz(K) edges. The edge energy of pristine zz(K) edge (the z6 configuration) is 4.11 eV/nm, which is the highest in all our explorations. Upon energy relaxation, the z6 configuration is able to transform into the (2×1) reconstructed z7 configuration, which is similar to the previously reported graphene zz(K) edge[9, 12]. However, different from the high instability of graphene zz(K) edge, the edge energy of the reconstructed z7 configuration is only 1.89 eV/nm, which is 0.47 eV lower than the (2×1) reconstructed z1 configuration (Fig. 1c). In addition, we also examine three other reconstructed edges of the zz(K) edges, that is, the z8, z9 and z10 as shown in Fig. 1b. The z8 configuration can be obtained by bending the edge atoms to form a small tube. The z9 configuration can be obtained by rolling up edge atoms to form a large tube. By opening up the nanotube in the z9 configuration, the z10 configuration can be obtained. Our calculations also show that the edge energy of the z8, z9 and z10 configurations is 0.97 eV/nm, 1.09 eV/nm and 0.53 eV/nm higher than that of the z7 configuration, respectively. Therefore, the z7 configuration is the most stable zz(K)-type edge. However, its edge energy is still 0.37 eV/nm (19.6%) higher than the z5 edge. Hence, our calculations suggest that the z5 configuration is the most energetically stable, and thus should be the dominated edge in terms of thermal stability. In a recent experiment,[34] it was found that phosphorene zigzag edge is semiconducting with a remarkable band gap, which is a clear evidence of edge reconstruction, and the z8 configuration was employed to explain the observed semiconducting characteristic. Thus, the realization of z5 edge is highly probable as its edge energy is 1.35 eV/nm lower than that of z8 edge. More interesting, as will be shown later, z5 edge is also semiconducting with a band gap of 1.23 eV.

It is noted that only free-standing phosphorene nanoribbons are considered in the present work. However, previous study showed that substrate may play a crucial role for the stability of phosphorene nanoflake. It was found that the edge of phosphorene nanoflake can roll up easily when it is in vacuum or on a substrate with a weak van der Waals (vdW) interaction, such as *h-BN* substrate[35]. On the other hand, if the interaction between phosphorene nanoflake and substrate is strong, for example, 5 times of that between phosphorene and *h-BN*, the energy barrier for rolling up can be increased significantly, and as a result, the formation of the tube-liked edge can be suppressed. Therefore, the reconstruction of tube-liked edge is expected only when phosphorene is on a substrate with a weak interaction, such as *h-BN*.

**Transition pathway and energy barrier**

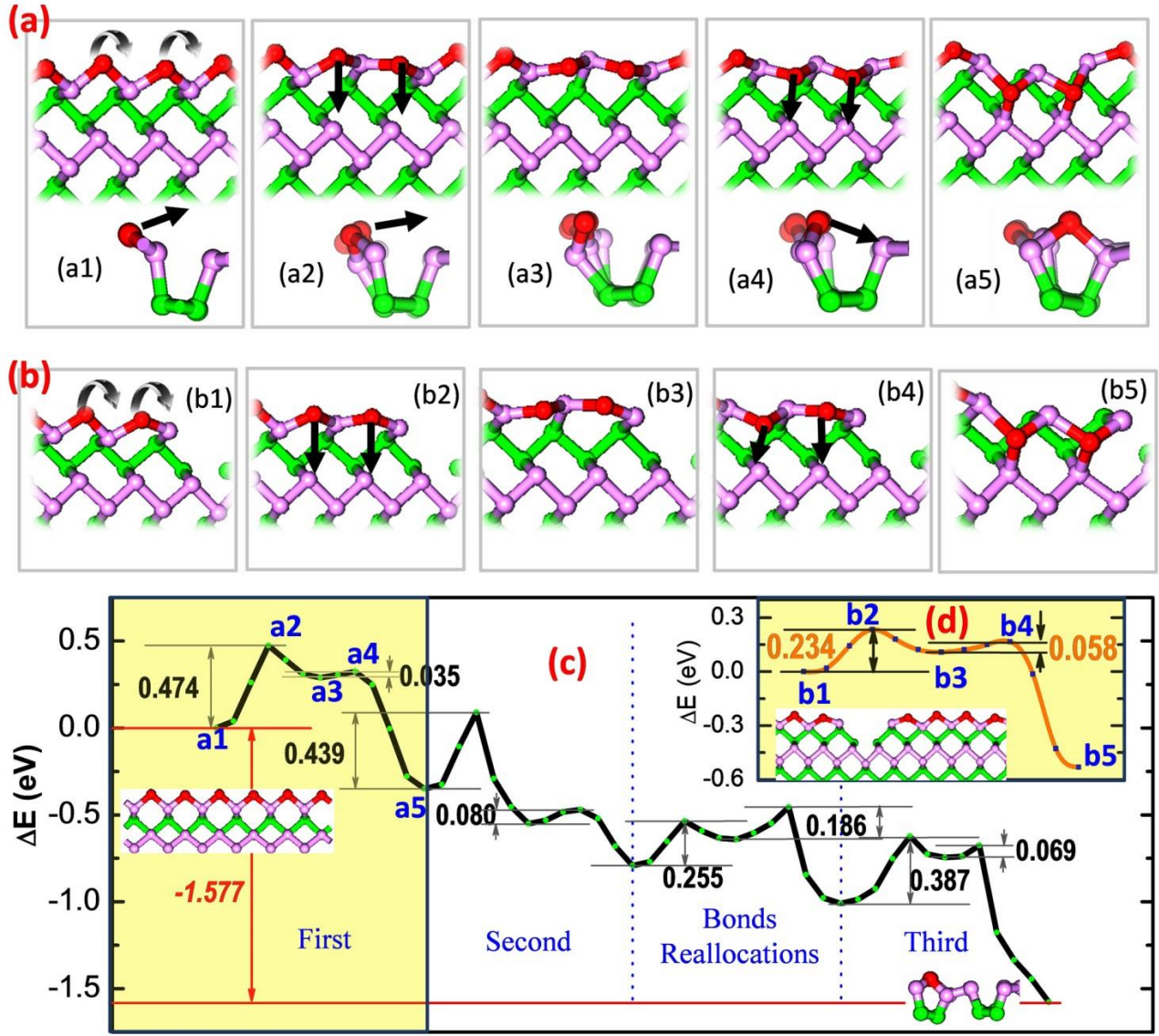

**Fig. 2** Reconstruction process from the pristine z1 configuration to the tubed z5 configuration. (a) a series of structure transformation in the reconstruction of the flat zz edge, (b) the consecutive transition structures for the zz edge with a kink, (c) the transition pathway and related energy changes from the flat z1 edge to the tubed z5 edge, and (d) the transition pathway and related energy changes starting from a kink site of the zz configuration.

In the above analysis, we have revealed surprisingly that the zz edges of PNRs can be terminated by nanotubes. In the following, we examine the transition pathway and related energy barriers from the z1 to z5 edge configurations (Fig. 2a) using 6-unit PNRs supercell along the edge. The change in the total energy was plotted in Fig. 2c, and the transitions and intermediate structures for the first two edge atoms were shown in Fig. 2a (a1-a5 in the yellow zone of Fig. 2c, and also the whole movie-S2 in ESI).

Overall, the z5 configuration is about 1.577 eV lower than the initial z1 configuration in the supercell, implying a strong driving force for the evolution. In the structural optimization of cNEB calculation, we find that when one P atom starts to roll up, it will bring the adjacent atom to move together. This process, however, has no direct influence on the next-nearest neighbour. Thus the formation of tube-liked edge is realized with a pair of adjacent edge atoms rolling up as a unit. The transition pathway and corresponding energy evolution for the first pair of edge atoms are shown in Fig. 2a and Fig. 2c, respectively. Initially, the two adjacent edge atoms roll up by overcoming an energy barrier of 0.474 eV (from a1 to a2) to reach the intermediate state (a3). Then, by overcoming a small barrier of 0.035 eV (from a3 to a4), the edge state a5 is formed, which is 0.35 eV lower than that of the a1 state. Subsequently, the adjacent two edge atoms begin to roll up and repeat the similar reconstruction behaviour as the first pair. The flipping barrier for the second pair is 0.439 eV, which is 0.035 eV lower than that of the first pair. For the third pair, the flipping barrier further decreases to 0.387 eV, implying that the subsequent flipping barrier may be monotonically decaying. It is important to note that after the rolling up of the first pair of P atoms, the energy barrier to form a tubed structure is only 0.035 eV. In contrast, to roll up the adjacent pair of P atoms, the energy barrier is 0.439 eV. Obviously, rolling up the adjacent pair of P atoms before the formation of a5 configuration of the first pair is not a kinetically favourable path. Therefore, the formation of z5 edge is through a stepwise manner: The first pair of atoms forms the closed a5 configuration, followed by the next pair and so on so forth.

During the roll-up of edge atoms, there are some consecutive P-P bond reallocations, corresponding to the change from z3 to z5 configuration. As shown in Fig. 2c, the reallocation barriers (0.255 eV and 0.186 eV) are much lower than the flipping barriers. This is in consistent with the easy reallocation of inner P-P bonds observed in previously *ab initio* molecular simulation of phosphorene nanoflake[35] and the low barrier of defect diffusion in phosphorene[36]. Overall, the highest energy barrier arises from the roll-up of edge atoms, especially for the first pair (0.474 eV). Thus we can estimate the average edge transition rate **R** by:

$$\boldsymbol{R} \sim (\frac{k_B T}{h}) \times exp(-\Delta E/k_B T) \quad (2)$$

where $T$ is the temperature, $k_B$ and $h$ are the Boltzmann and Plank constants, respectively. $\Delta E$=0.474 eV. At room temperature (300K), the average edge transition rate **R** is about $10^5$ Hz. Besides, at elevated 400 K, **R** reaches to $10^6$ Hz. At the thermal decomposition temperature ~400 °C,[37] the average edge transition is $10^9$ Hz, indicating that the edge transition will happen in just several nanoseconds at this temperature.

In general, an edge atomic structure may be not perfect. Our DFT calculation shows that the formation energy of a kink (the inset structure in Fig. 2d) along the edge is only 0.22 eV/atom. Therefore, the possibility to form a kink along the edge is about *exp(-0.22 eV/$k_B T$)*, which is ~$10^{-4}$ at 300 K. Thus, there should be a kink in the micrometre scale. In reality, the density of kink along the edge could be even higher.[34]

We then investigate the effect of kink on the edge reconstruction. Fig. 2b and 2d show the transition structures and energy barriers for the roll-up of the first pair of two adjacent edge atoms near the kink. Although the transition processes are very similar to those of the regular z1 edge, the highest flipping barrier is reduced to 0.234 eV, which is only half of the regular edge (0.474 eV). Therefore, it is expected that the tubed z5 configuration can be formed rapidly, and should be the dominant configuration for phosphorene in vacuum or on a substrate with a weak vdW interaction even at room temperature.

**Phase diagram analysis**

Similar to graphene[10], the structure and stability of phosphorene edge may be greatly affected when exposed to $H_2$ gas[19, 38]. In the following, we compare the stability of z1, z5 and hydrogenated zz configurations. The edge energy γ(H) of hydrogenated zz edge can be calculated by[10]:

$$\gamma(H) = \frac{1}{2L}(E_{ribb} - N_P E_P - \frac{N_H}{2} E_{H_2}) \quad (3)$$

where $E_{H_2}$ is the energy of a $H_2$ molecule obtained from DFT calculations. The γ(H) is about 0.18 eV/nm, which is slightly higher than the previous results since the DFT-D3 is adopted in our calculations. The change of Gibbs free energy associated with the related stability of hydrogenated edge with respect to the bare edge can be calculated by:

$$\Delta G = \gamma(H) - \frac{\rho_H}{2}\Delta\mu(H_2) \qquad (4)$$

where $\rho_H$ is the concentration of H atoms along the hydrogenated edge. At the absolute temperature T and under a certain partial pressure $P(H_2)$, the chemical potential of $H_2$ gas is:

$$\Delta\mu(H_2) = H^0(T) - H^0(0) - TS^0(T) + k_B T \ln\left[\frac{P(H_2)}{P_0}\right] \qquad (5)$$

where the $H^0(T)$ and $S^0(T)$ can be obtained from the NIST-JANAF thermochemical tables,[39] the reference pressure is $P_0$=1 bar.

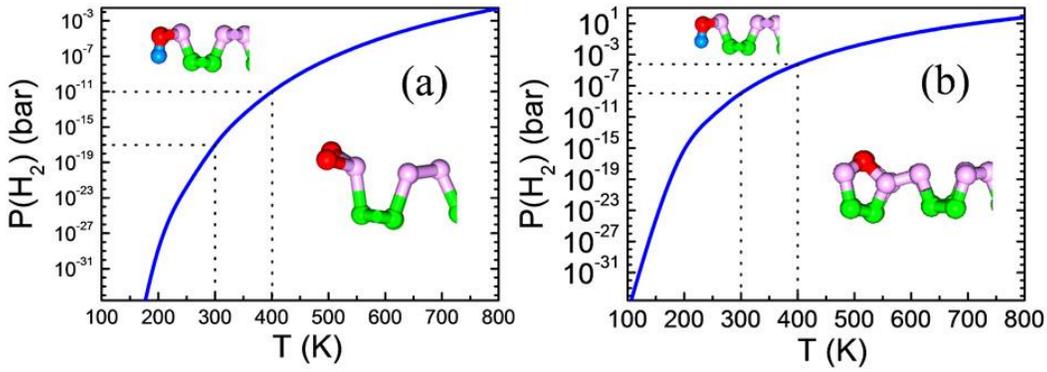

**Fig. 3** The phase diagram of the bare and hydrogenated zz edges under various temperatures and partial pressures of hydrogen gas. (a) the z1 edge vs. the zz-H edge; and (b) the tubed z5 edge vs. the zz-H edge.

Fig. 3 presents the phase diagram of the hydrogenated edge in comparison with the bare z1 edge (Fig. 3a) and the tubed z5 edge (Fig. 3b), respectively. It is seen that the bare z1 edge can only exist under very low $H_2$ pressure. At 300 (400) K, the z1 edge can only exist under an ultra-low $H_2$ pressure of $10^{-17}$ ($10^{-11}$) bar (Fig. 3a). At lower temperature, the transition hydrogen pressure should be even lower.

As mentioned above, the z1 edge without hydrogenation will rapidly transform into the tubed z5 edge at room temperature and above. After this reconstruction, the chemical stability of the z5 edge is significantly enhanced. At room temperature, the transition pressure $P(H_2)$ from the tubed z5 configuration to the hydrogenated zz configuration is ~$10^{-8}$ bar, which is close to the $H_2$ partial pressure in air (~$10^{-7}$ bar). At elevated temperature, the transition pressure $P(H_2)$ is above the $H_2$ partial pressure in air. For example, at 400 K, the transition pressure $P(H_2)$ is ~$10^{-4}$ bar, which is three orders higher than the H$_2$ partial pressure in air at the same temperature. Usually, the production and fabrication of phosphorene are under high vacuum condition,[38, 40, 41] where the hydrogen pressure is usually much lower than the partial pressure in air. Therefore, the most stable phosphorene edge should be the tubed z5 edge in the common practical condition.

Furthermore, it was shown that when phosphorene is directly exposed to the air, surface of phosphorene can be easily oxidized, leading to the formation of complex P$_x$O$_y$ oxides[42] and the degradation of phosphorene[43]. Although the oxidation of phosphorene sheet has been explored, the effect of oxygen molecules on the edge reconstruction and stability remains unexplored and deserves further investigation.

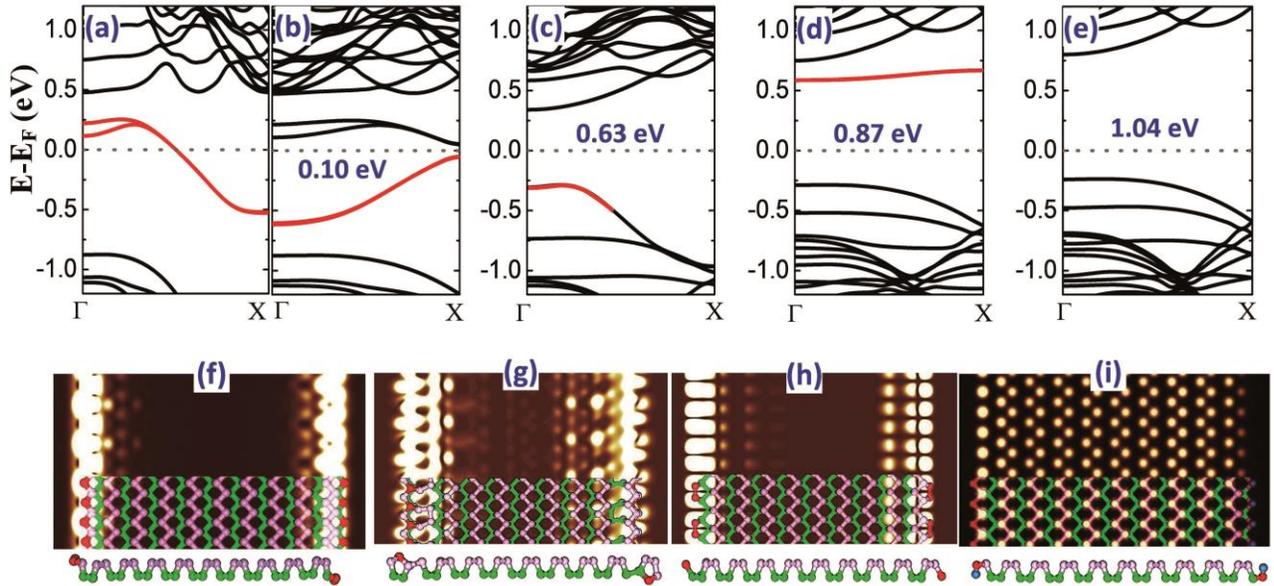

**Fig. 4** The band structures of PNRs with different edges. (a) the pristine zz edge; (b) the z1 edge; (c) the z5 edge terminated with a nanotube; (d) the z7 edge; and (e) the hydrogenated edge. The simulated STM images of the z1 edge at -0.7 V bias (f); the z5 edge at -0.5 V bias (g), the z7 edge at +0.7 V bias (h), and the hydrogenated zz edge at -1.5 V bias (i). The red lines denote the band due to the edge contribution.

**Electronic properties of various edges**

Next we investigate the electronic properties of the pristine zz, (2×1) reconstructed z1, tubed z5, zz(K)-type z7 and hydrogenated zz edges using wide asymmetric PNRs. It should be noted that to suppress the reconstruction, the pristine zz edge includes only one periodic unit along the edge. The calculation results are shown in Fig. 4. It is seen that the band structure of the pristine zz edge exhibits a clearly metallic character (Fig. 4a), in agreement with previous calculations.[19-21,25] After (2×1) reconstruction, the PNR become a semiconductor with a narrow band gap of 0.10 eV (Fig. 4b).

It is seen that the PNR with the tubed z5 edges possesses a band gap of 0.63 eV (Fig. 4c), which is 0.53 eV larger than the reconstructed z1 edge. It is also seen that the edge state in band structure is close to the valence band. We also employed HSE06 functional[44] to find more accurate band structure; it is found that both standard PBE and HSE06 can describe the band arrangement order reasonably [see Fig. 4c and Fig. 5]. But the band gap with HSE06 functional is 1.23 eV [Fig. 5], which is about twice than that from standard PBE calculation.

We find that the z7 configuration also has a large band gap of 0.87 eV (Fig. 4d). However, the edge state in the band gap of the z7 configuration is close to the conduction band. For the hydrogenated phosphorene edge, the band gap is 1.04 eV, which is nearly the same as the perfect phosphorene monolayer calculated using standard GGA. In addition, there is no edge state in the band gap (Fig. 4e).

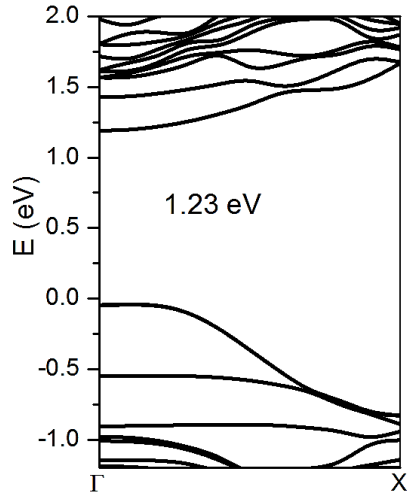

**Fig.5** The HSE06 calculated band structures of PNRs with z5 edge terminated by a nanotube. The band gap increases to 1.23 eV, which is about twice that from the standard PBE calculation. It is noted that the HSE06 calculated band arrangement is similar to PBE calculations.

Currently, it is still a significant challenge to experimentally resolve the fine edge structure due to the considerable height fluctuation near the edge. Besides, the high resolution signals can only be obtained at a small bias window related to the edge state position in the gap. To facilitate experimental observations, we simulate STM images of these typical edges. We find that there are large differences in the edge signals among the z1 (Fig. 4f), the tubed z5 edge (Fig. 4g) and the z7 edge (Fig. 4h), under corresponding bias according to the red lines in their band structures, i.e. -0.7 V for z1, -0.5 V for z5 and +0.7 V for z7 edge. Note that even though we have used -1.5 V bias for hydrogenated edge (Fig. 4i), no significant edge state is found in the simulated STM images.

**Conclusions**

In summary, we have investigated various possible reconstructions for pristine phosphorene zz and zz(K) edges. Surprisingly, the most stable edge is found to be the one terminated by a nanotube configuration. Of practical interest is that this edge configuration can easily occur at room temperature, and is stable against hydrogenation even under common production and fabrication process of phosphorene. Furthermore, our calculations show that tube-terminated edge has a band gap of 1.23 eV with obvious edge state close to the valence band. Finally, we have also provided the simulated STM images to facilitate experimental observation. It is expected that the compelling evidences presented here for the presence of the tube-terminated zigzag edge will stimulate more studies in finding other novel edge types, and further exploring their practical applications.

**Acknowledgements**

This work was supported in part by a grant from the Science and Engineering Research Council (152-70-00017). The authors gratefully acknowledge the financial support from the A*STAR, Singapore and the use of computing resources at the A*STAR Computational Resource Centre, Singapore.

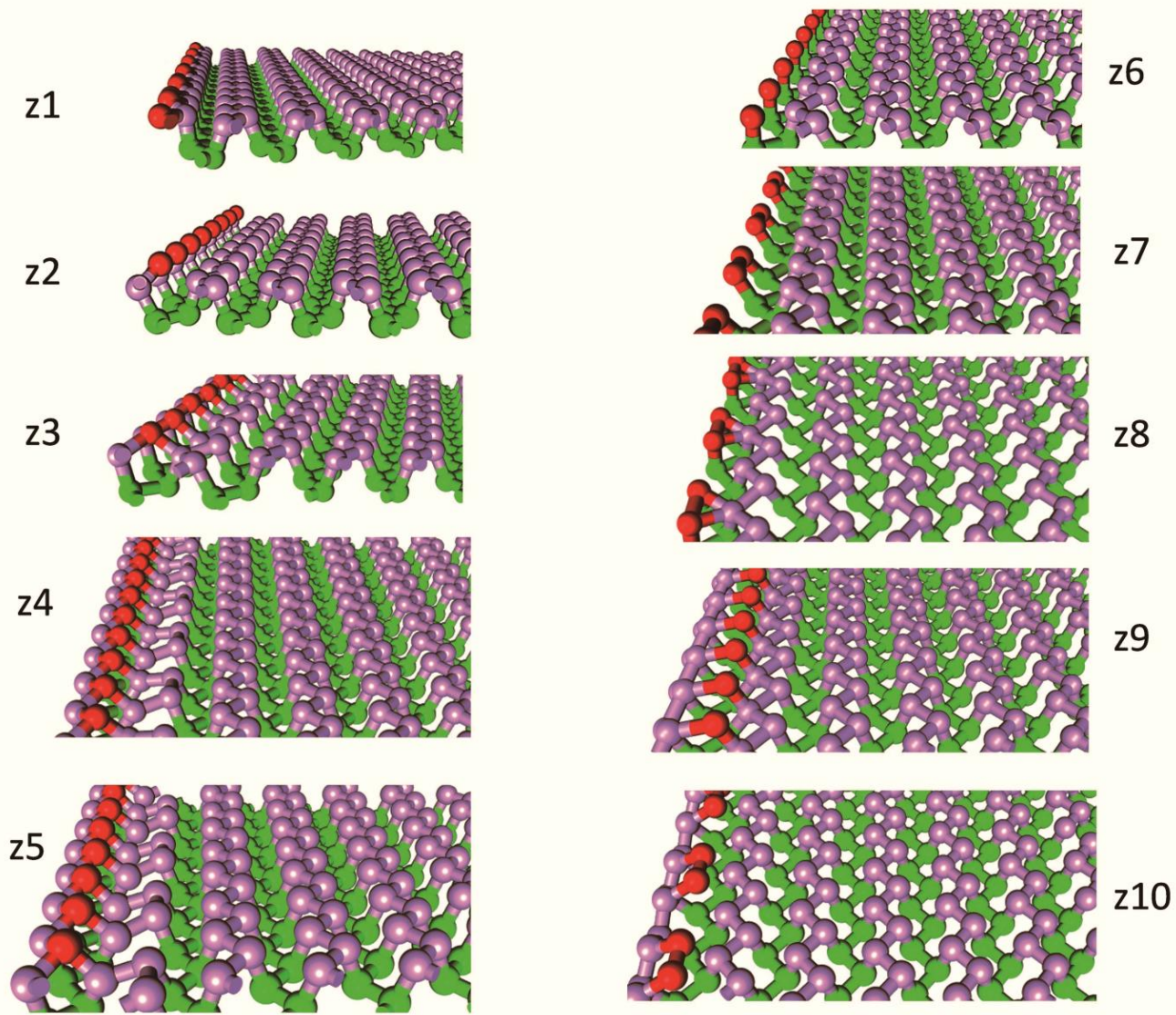

**Fig. S4** The perspective views of various edge structures explored in the present study.